\documentclass[traditabstract]{aa}
\usepackage{graphicx}
\usepackage{txfonts}

\def\tex {\ifmmode{{T}_{\rm ex}}\else{$T_{\rm ex}$}\fi}
\def\tmb {\ifmmode{{T}_{\rm mb}}\else{$T_{\rm mb}$}\fi}
\def\ci     {\ifmmode{{\rm C}{\rm \small I}}\else{C\ts {\scriptsize I}}\fi}
\def\hi     {\ifmmode{{\rm H}{\rm \small I}}\else{H\ts {\scriptsize I}}\fi}
\def\hh     {\ifmmode{{\rm H}_2}\else{H$_2$}\fi}

\def\ts     {\thinspace}
\def\kms    {\ifmmode{{\rm \ts km\ts s}^{-1}}\else{\ts km\ts s$^{-1}$}\fi}
\def\msol   {\ifmmode{{\rm M}_{\odot}}\else{M$_{\odot}$}\fi}
\def\lsol   {\ifmmode{{\rm L}_{\odot}}\else{L$_{\odot}$}\fi}
\def\zsol   {\ifmmode{{\rm Z}_{\odot}}\else{Z$_{\odot}$}\fi}

\usepackage{color}
\begin{document}

\title{Bulge formation in disk galaxies with MOND}

\author{F. Combes \inst{1}
           }
\offprints{F. Combes}
\institute{Observatoire de Paris, LERMA (CNRS:UMR8112), 61 Av. de l'Observatoire, F-75014, Paris, France
\email{francoise.combes@obspm.fr}
              }
\date{Received September 15, 2014/ Accepted September 25, 2014}

\titlerunning{Bulge formation in disk galaxies with MOND}
\authorrunning{F. Combes}

\abstract{The formation of galaxies and their various components can be stringent tests
of dark matter models and of gravity theories. In the standard cold dark matter (CDM) model,
spheroids are formed through mergers in a strongly hierarchical scenario and also in the early
universe through dynamical friction in clumpy galaxies. More secularly, pseudo-bulges 
are formed by the inner vertical resonance with bars. The high efficiency of 
bulge formation is in tension
with observations in the local universe of a large amount of  bulgeless spiral galaxies.
In the present work, the formation of bulges in very gas-rich galaxies, such as those
in the early universe, is studied in Milgrom's MOdified Newtonian 
Dynamics (MOND) through multigrid simulations of the nonlinear
gravity, including gas dissipation, star formation, and feedback. 
 Clumpy disks are rapidly formed, as in the equivalent Newtonian systems. 
However, the dynamical friction is not as efficient in the absence of dark matter
halos, and the clumps have no time to coalesce into the center to form bulges
before they are eroded by stellar feedback and shear forces.
Previous work has established that mergers are less frequent in MOND, 
and classical bulges are expected less massive. We now
show that gas-rich clumpy galaxies in the early universe do not form bulges.
 Disks with a low bulge fraction, which is compatible with the observations,
 are therefore a natural result in MOND.
 Since pseudo-bulges are formed by bars with a similar rate to those in the Newtonian equivalent systems,
it can be expected that the contribution of pseudo-bulges is significantly higher in MOND.
}
\maketitle


\section{Introduction}

Many problems in cosmology that are related either to the missing mass in galaxies and clusters 
or to the formation of large-scale structures and galaxies  are solved by the 
existence of a dark sector, dark matter, and dark energy, dominating the content
of the universe (e.g., Blumenthal et al. 1984, Peebles \& Ratra 2003). The standard hypothesis
 for dark matter is the cold and massive elementary particle (CDM) coming from supersymmetry
(Bertone et al 2005), although there is also a recent revival of warm dark matter models that are
able to reproduce the power spectrum of structures better at galactic scales and below  
(Bode et al. 2001). All dark matter models, however, encounter persistent problems at galactic scales,
such as the predicted cusps and abundance of dark matter in galaxies that is too high (e.g.,
Wang \& White 2009, Kennedy et al. 2014, Boylan-Kolchin et al. 2011, 2012).

An alternative to dark matter models that work in Einstein gravity is to assume
that a modification of the gravity could account for the apparent dark sector and, in particular,
the MOdified Newtonian Dynamics (MOND) proposed by Milgrom (1983), which 
accounts for the missing mass on galaxy scales very well (see the review by Sanders \& McGaugh
2002 and Famaey \& McGaugh 2012). The fundamental idea is based on the observation
that mass discrepancies mainly occur when the acceleration falls below the critical
value of a$_0 \sim$ 2 10$^{-10}$ m/s$^2$. At low acceleration, the gravitational force
varies as 1/r instead of 1/r$^2$. The outer parts of giant galaxies and dwarf galaxies
should fall entirely into this regime.

In standard dark matter models, bulges in galaxies are thought to be formed mainly
through major or minor mergers (e.g., Toomre 1977, Barnes \& Hernquist 1991,
Naab \& Burkert 2003, Bournaud et al. 2005, 2007a). The time scale of a merger
is short, on the order of 10\% of Hubble time or less, because of the existence of
extended and massive dark halos, which take the orbital angular momentum of baryons away
(Barnes 1988). In modified gravity models and in the absence of halos, the spiral-in of companion 
galaxies occur on a much larger time scale, and 
the frequency of mergers is low (Tiret \& Combes 2008b). The prediction of the number
and mass of bulges in spiral galaxies could then be a discriminant between models.
 Another way to form bulges occurs in the early universe, when gas-rich disks are unstable
to collapsing into massive clumps, which spiral in and form a bulge and thick disk 
(e.g., Noguchi 1999, Bournaud et al. 2007b). Dynamical friction is thought to bring the
clumps inward, faster than their destruction through star formation feedback, and this
could also constrain dark matter models. 

There is a third main dynamical process able to form bulges in spiral galaxies,
the vertical resonance with a stellar bar (Combes et al. 1990, Kormendy \& Kennicutt 2004).
These are called pseudo-bulges, since they show more rotation, more flattening, and a 
radial distribution closer to an exponential than do ``classical" bulges. Bars form at 
comparable frequency in MOND and Newtonian models with dark matter (Tiret \& Combes,
2007, 2008a), and pseudo-bulges are expected to form as frequently in the two models.

The goal of the present article is to compare the
formation of bulges in clumpy disks in the frame of dark matter and MOND models, to
probe early bulge formation.
The methods for the simulations are described in Section \ref{simul}, and the galaxy 
models used in Section \ref{model}. The results are first presented
for dynamical friction in dissipationless models (Section \ref{stars}),
and then all the phenomena of gas-rich disks forming clumpy galaxies 
with star formation and feedback are included  in Section \ref{gas},
and discussed in Section \ref{disc}.
The conclusions of bulge formation in the different models are then summarized in 
Section \ref{summary}.

\section{Simulation methods}
\label{simul}

Since the MOND equations are nonlinear, it is not possible to use
normal Poisson solvers, such as the Tree code, but instead the full 
differential equations should be solved on a grid. 
This is very efficiently done with the multigrid algorithm 
(cf. Numerical Recipes, Press et al., 1992, Brada \& Milgrom 1999, Tiret \& Combes 2007,
hereafer TC07).
 Both the MOND simulations and the Newtonian gravity comparison runs
were carried out with the same multigrid method. 
For the modified gravity runs, the self-consistent Lagrangian theory AQUAL,
developed by Bekenstein \& Milgrom (1984) was used, 
where the Poisson equation is written as

\begin{equation}
\label{eq:mondaqual}
\nabla [ \mu(|\nabla\Phi|/a_0) \nabla\Phi ] = 4 \pi G \rho,
\end{equation}

\noindent where $\mu(x)$ is the interpolation function that is equal to unity at large $x$
(Newtonian regime), and it tends to $x$ when $x <<$ 1 in the MOND regime.
 Various functions have been used in the literature, and for fitting 
rotation curves of spiral galaxies, the best one appears to be the standard function
 $\mu(x)= x/(1+x^2)^{1/2}$ (Sanders \& McGaugh 2002), while for the Milky Way, the 
best fit is obtained with the so-called "simple" function
 $\mu(x)= x/(1+x)$ (Famaey \& Binney 2005). The values of a$_0$ range between
1 and 2 10$^{-10}$ m/s$^2$, according to the $\mu$ function used: with the simple function,
the MOND regime is reached with an acceleration about one order of magnitude
greater than with the standard function (e.g., Tiret \& Combes 2008a). There are
other uncertain parameters that could influence the choice of the coupled
parameters ($\mu$, a$_0$), such as the unknown fraction of dark baryons 
in galaxy disks (Tiret \& Combes 2009) or the impact of the external
field effect (Zhao \& Famaey 2006).

The multigrid method developed in the present work has been described
at length and tested in TC07, in particular the problem of boundary conditions. Equation (\ref{eq:mondaqual}) has been discretized similarly, and 
the Gauss-Seidel relaxation with red and black ordering was followed.
If the convergence is very quick with the linear Newtonian Poisson equation,
the analogous MOND equations take a much longer CPU time. Then to be able to run
many more simulations in a given time, I used the quasi linear formulation of MOND,
call QUMOND developed by Milgrom (2010).

This formulation has the advantage of being much easier to solve, since
the interpolation function of a/a$_0$ has been transferred to its inverse equivalent
as a function of a$_N$/a$_0$, with a$_N$ the Newtonian acceleration.
It is a non-relativistic complete theory, derivable from an action,
but the nonlinearity can be reduced to solve the linear Poisson equation several times.
Indeed, the equation to solve is the following:

\begin{equation}
\label{eq:qumond}
\nabla^2 \Phi = \nabla [ \nu(|\nabla\Phi_N|/a_0) \nabla\Phi_N ] 
\end{equation}

\noindent where $\Phi$ is the Milgromian potential and $\Phi_N$ the Newtonian one.
Since the normal Poisson equation links $\Phi_N$ and the density $\rho$
by the relation 
$\nabla^2 \Phi_N = 4 \pi G \rho$,
equation (\ref{eq:qumond}) can be written as the function of two densities:
$$
\nabla^2 \Phi =  4 \pi G \rho_b + 4 \pi G \rho_{ph}
$$
\noindent where $\rho_b$ is the
 baryonic density, and $\rho_{ph}$ the so-called phantom density,
which is the equivalent of the missing mass within the Newtonian gravity.
Examples of the phantom density distribution are given by L\"ughausen et al. (2013),
it can sometimes (rarely) be negative.
The phantom density can be computed by

\begin{equation}
\label{eq:phantom}
\rho_{ph}= \frac{1}{ 4 \pi G}  \nabla [(\nu -1) \nabla\Phi_N ]
.\end{equation}
  
The computation of the Milgromian potential therefore can be carried out
in two linear steps. First, the Newtonian potential $\Phi_N$ is computed on the grid
from the baryonic density $\rho_b$, and the phantom density $\rho_{ph}$
is derived on the same grid, through discretization of equation (\ref{eq:phantom})
(see, e.g., Famaey \& McGaugh 2012), in a similar manner to equation  
(\ref{eq:mondaqual}). Then the linear Newtonian Poisson solver is used
again, this time with $\rho_b + \rho_{ph}$.
The interpolation function $\nu (a_N/a_0)$ corresponding to the standard
$\mu$ function can easily be computed as
$\nu(y)=\sqrt{0.5+0.5\sqrt{1+4/y^2}}$. The two MOND formulations 
AQUAL and QUMOND are 
not completely equivalent, since their forces differ by a curl field 
(Zhao \& Famaey 2010).
 The two are very close in symmetric systems, however, and in what follows,
the results from the two formulations were checked for similarity.

\bigskip

\begin{figure}[ht]
\centerline{
\includegraphics[angle=-90,width=4cm]{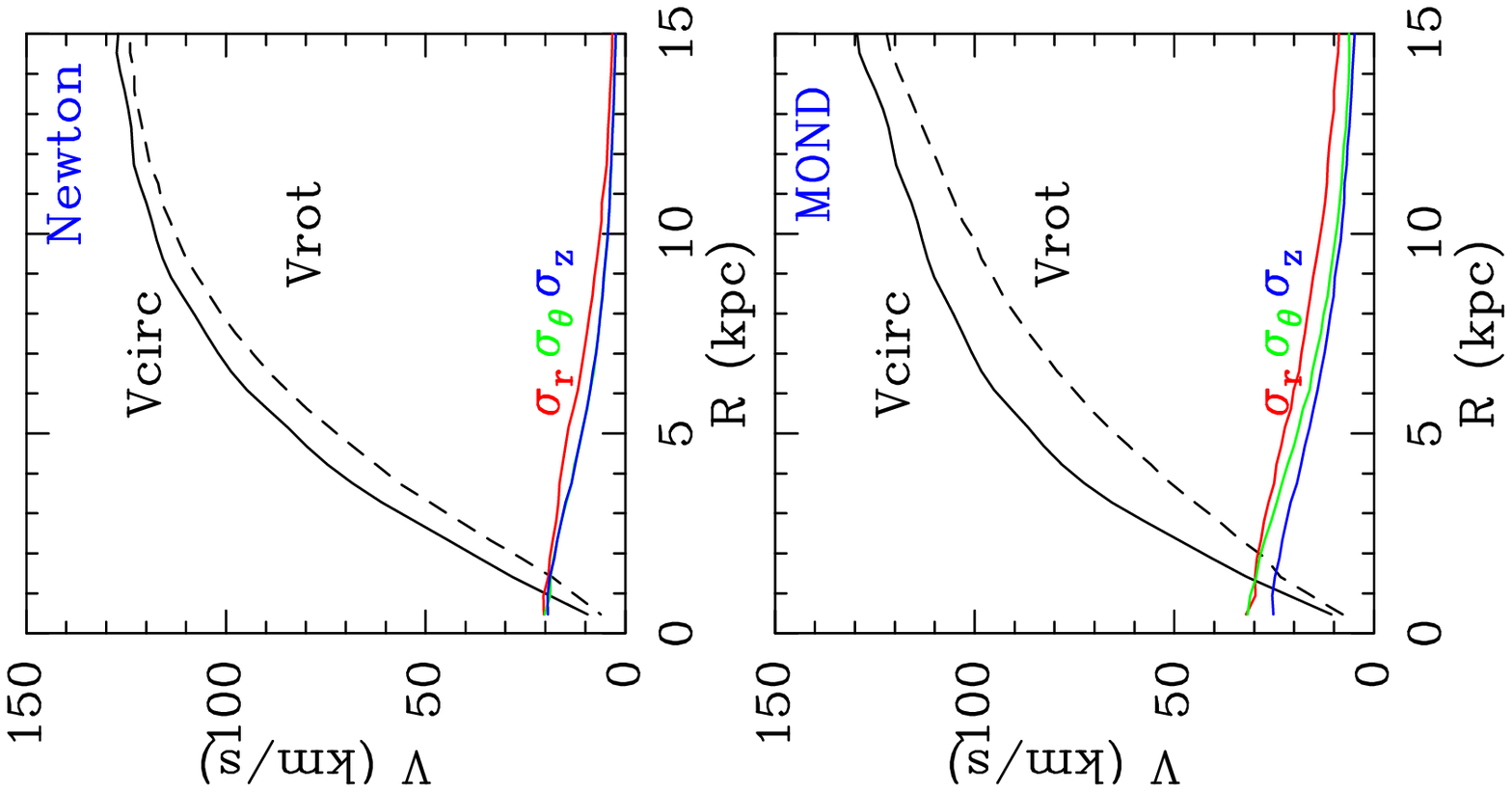}
\includegraphics[angle=-90,width=4cm]{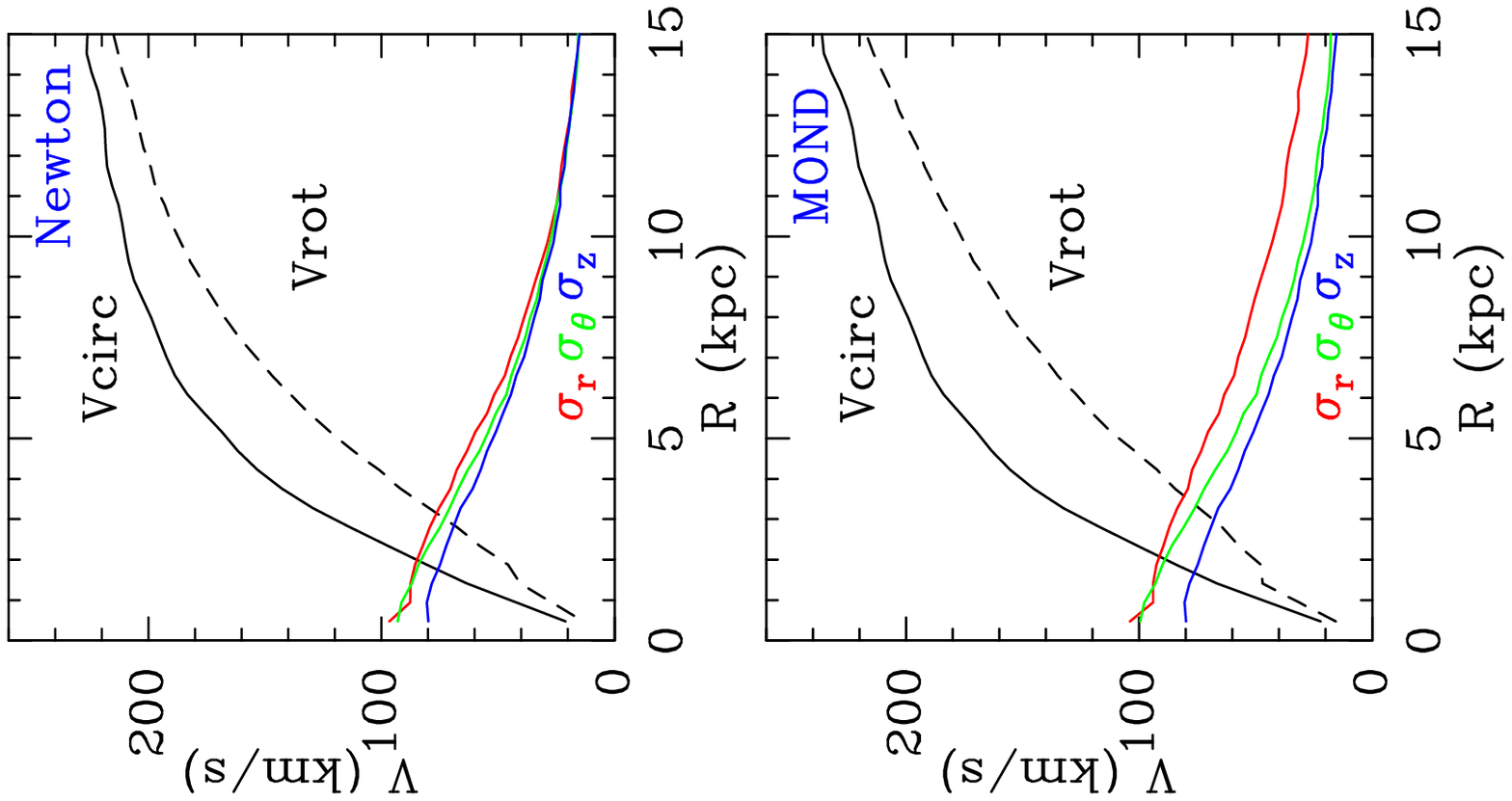}}
\caption{{\bf Left}: Circular and rotational velocities, together
with the velocity dispersions for the stellar component
in the Newtonian +DM  (top) and MOND (bottom) 
initial galaxy models with stars and gas.
{\bf Right}: Same for the giant galaxy model.}
\label{fig:rotc}
\end{figure}

%
\begin{table*}[ht]
\begin{center}
      \caption[]{Simulation models} 
\label{tab:simul}
\begin{tabular}{llrrrrrrrrrrr}
\hline
Model & M$_*$ & r$_*$  & z$_*$ & M$_g$ & r$_g$ & z$_g$ & $\beta_r$ & $\beta_t$ & C$_{star}$ & $\alpha$&  M$_{H}^{(1)}$ & r$_{H}$ \\
     &  10$^9$\msol & kpc & kpc &10$^9$\msol & kpc& kpc &     &     &    &       & 10$^{10}$\msol  &  kpc \\
\hline
dwarf$^{(*)}$ & 5.68 & 5.  &  1.  &  0. & -  & - &  -  & -  &  - &  -  &  7.5  & 12  \\
dwarf & 2.84 & 5.  &  1.  &  2.84 & 5.  & 0.5 &  0.5  & 0.5  &  2 10$^{-4}$ &  1.4  &  7.5  & 12  \\
giant$^{(*)}$ & 56.8 &  5. &  1.  &  0. & -   & -  &  -   & -   &   -  &  -   &  17.6  & 16  \\
giant & 28.4 &  5. &  1.  &  28.4 & 5.  & 0.5 &  0.5  & 0.5  & 2 10$^{-4}$ &  1.4  &  17.6  & 16  \\
\hline
\end{tabular}
\\ $^{(*)}$ These runs are purely stellar
\\ $^{(1)}$ Dark halo mass of the equivalent Newtonian system 
\\ The depletion time scale is 5 Gyr, for the homogeneous gas disk
\end{center}
\end{table*}

The grid used in the present work is 257$^3$ in size. The simulation box covers a 
$60\ \mathrm{kpc}$ cube, and the spatial resolution is about $230\ \mathrm{pc}$.
 The rest of the simulation code works like a particle-mesh (PM) code.
Up to 0.75 million particles are initially distributed on the grid.
 The density is computed on the grid by a cloud-in-cell (CIC) algorithm,
and once the potential is computed, particles are advanced using a leap-frog 
method. For the purpose of distinguishing the various physical mechanisms,
simulations with only stars (and dark matter in the Newtonian equivalent systems) were run, followed by simulations that included gas, star formation, 
and feedback. The gas dissipation was taken into account with a sticky-particle
code, as described in Tiret \& Combes (2008a). The rebound parameters
were varied from 0.4 to 1, which essentially affected the amount of 
dissipation and the time scales of clump formation. The runs presented in the
following are all with $\beta_r = \beta_t = 0.5$. The variety of results obtained
with a changing gas dissipation does not fundamentally change the conclusions.

Star formation follows a Schmidt law; i.e., the star formation rate (SFR) is proportional
to a power $n$ of the local volumic density with a density threshold. 
The exponent $n$ has been taken equal
to 1.4 (Kennicutt 1998). This nonlinear relation means that gas clumps will primarily be the site of star formation. The proportionality factor is selected
to give a depletion time scale for a homogeneous gas disk of 5 Gyr in its initial
state. When clumps form, the consumption time scale is considerably reduced
owing to the higher star formation efficiency in dense clumps.
A continuous mass loss from stars into the gas is taken into account 
as in Jungwiert et al (2001). The mass loss by stars
is distributed through the gas on neighboring particles, with a velocity dispersion 
to schematize the feedback energy.

\begin{figure}[ht]
\centerline{
\includegraphics[angle=-90,width=4cm]{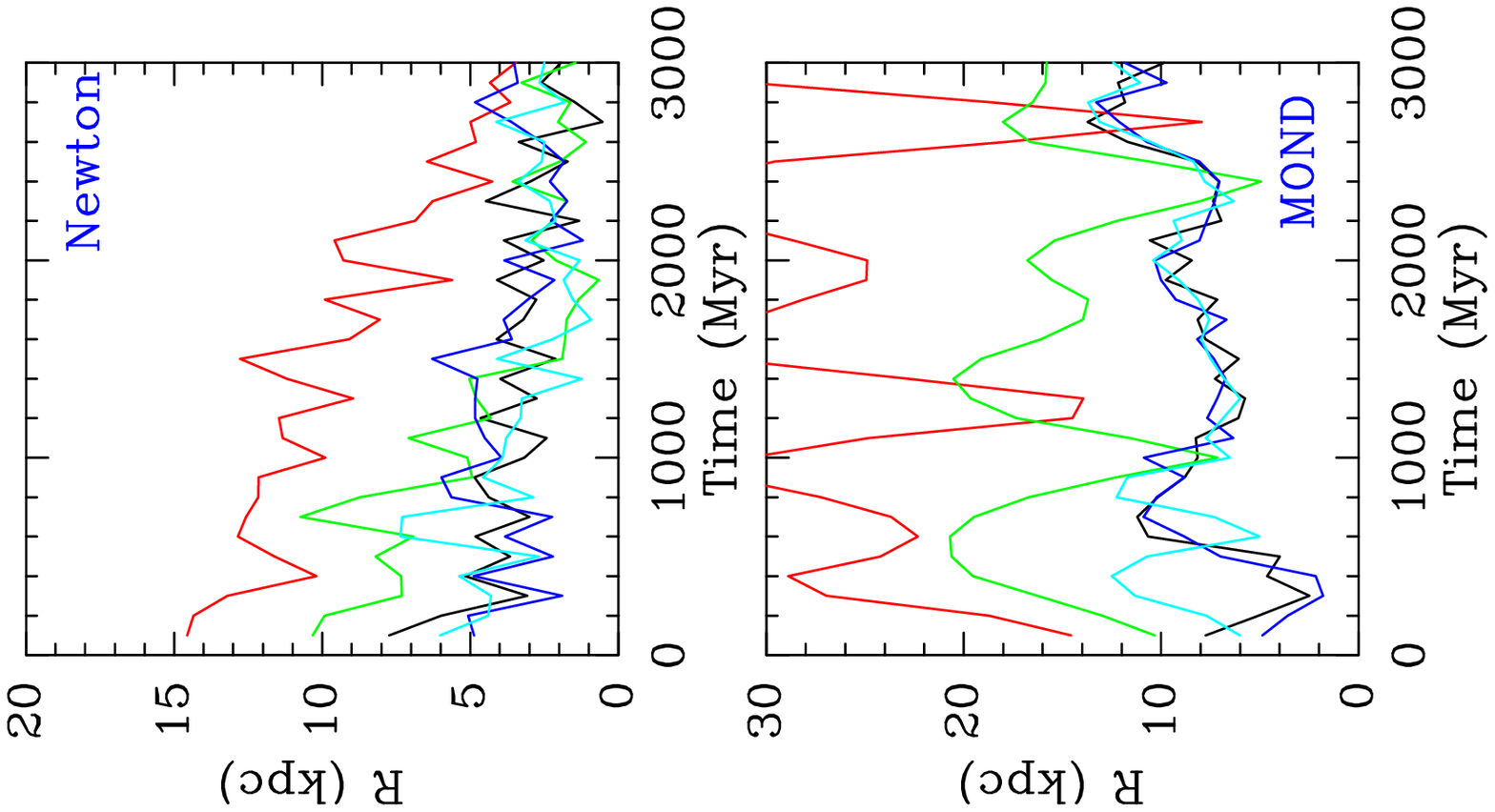}
\includegraphics[angle=-90,width=4cm]{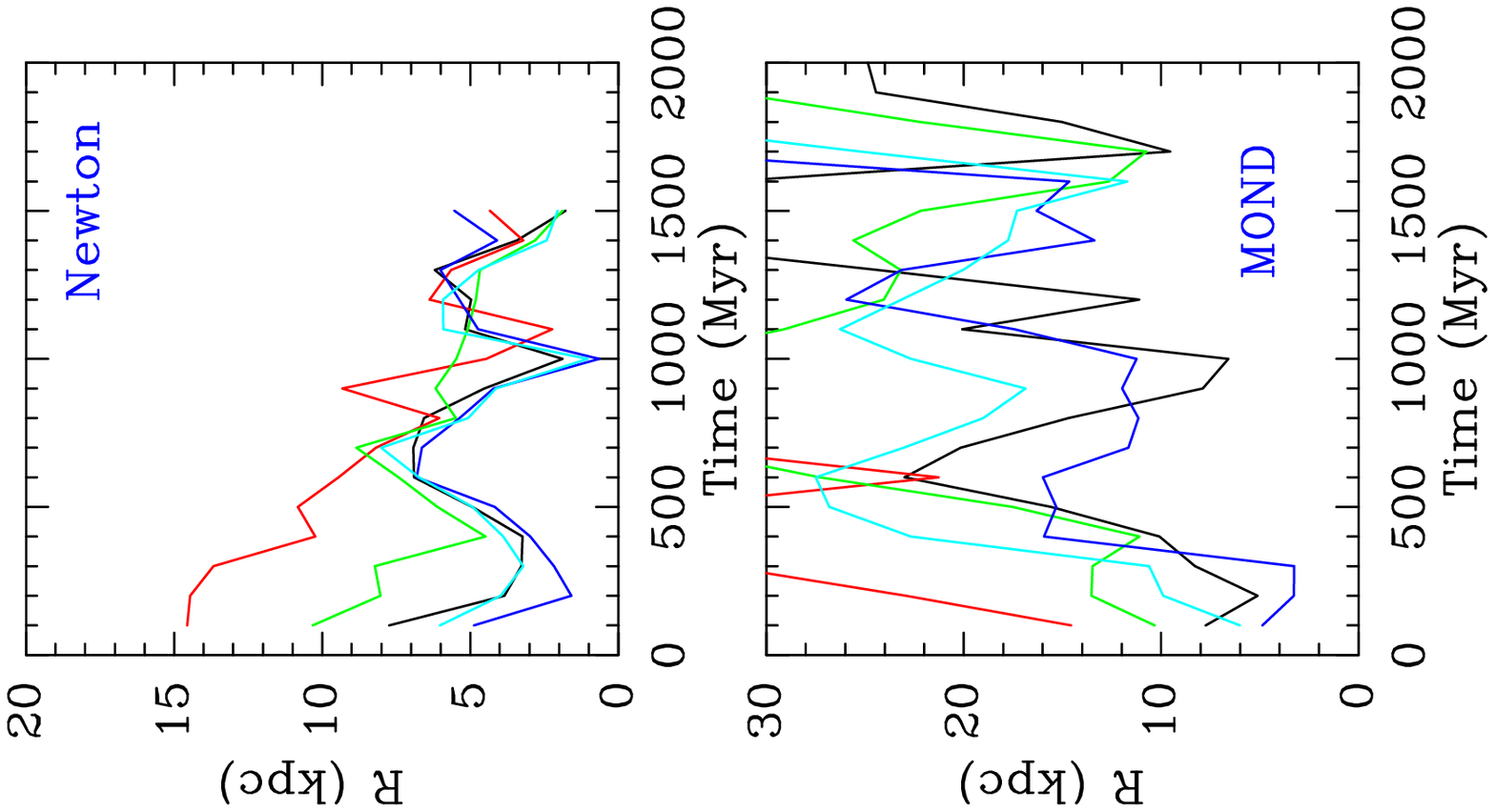}}
\caption{{\bf Left}: Decay of the 5 clumps containing globally
25\% of the baryonic disk as a function of time,
in the Newtonian +DM  (top) and MOND (bottom), for 
the purely stellar galaxy models.
{\bf Right}: Same for the giant galaxy model.}
\label{fig:decay}
\end{figure}

\section{The model galaxies}
\label{model}

The goal of the present work is to study the formation of bulges
in the early universe, therefore the initial conditions are gas-rich galaxies
with a gas fraction of 50\% of the baryonic mass and no bulge to start with.

The stellar disk follows a Miyamoto-Nagai density profile:

\begin{eqnarray}\label{stdisk}
\rho_{*}(R,Z)&=&\left(\frac{{z_{*}}^2 M_{*}}{4 \pi}\right)\times\nonumber\\&&\frac{r_{*} R^2+(r_{*}+3\sqrt{Z^2+{z_{*}}^2})\left(r_{*}+\sqrt{Z^2+{z_{*}}^2}\right)^2}
{\left[R^2+\left(r_{*}+\sqrt{Z^2+{z_{*}}^2}\right)^2\right]^{5/2}\left(Z^2+{z_*}^2\right)^{3/2}}
\end{eqnarray}

\noindent with mass $M_{*}$ and radial and vertical scale lengths 
$r_{*}$ and $z_{*}$, respectively.
The gas disk follows the same distribution with characteristic parameters
 $M_{g}$, $r_{g}$, and $z_{g}$, respectively.
All values of parameters are displayed in Table \ref{tab:simul}.
The simulation resolution ($230$ pc) is barely enough to study the vertical 
structure of the gas disk ($500$ pc).
All components are truncated at three times the value of their characteristic scale,
therefore the extent of the disk is initially 15 kpc in radius.

In the equivalent Newtonian galaxies, the dark halo 
is modeled as a Plummer sphere with
characteristic mass $M_H$ and characteristic radius 
$R_H$. Its density is given by the following analytical
formula:

\begin{equation}\label{halo}
\rho_{H}(R)=\left(\frac{3M_{H}}{4\pi {r_{H}}^3}\right)\left(1+\frac{R^2}{{r_{H}}^2}\right)^{-5/2}
.\end{equation}
The values of these parameters M$_H$ and r$_H$ are selected so that 
the Newtonian equivalent model has the same rotation curve as the disk
in the modified dynamics.
The dark matter halo is truncated at 21 kpc, and the halo masses
indicated in Table \ref{tab:simul} are relevant to this radius.

The stability of these galaxy models is controlled initially through a
Toomre parameter for the stellar disk of $Q_{\mathrm{star}}=1.5$ and 
for the gas component, $Q_{\mathrm{gas}}=1.1$.
The initial rotation curves of the model galaxies are shown in Fig.\ref{fig:rotc}.
 To avoid artificial initial relaxation of the disks, particles are
advanced during a typical rotation period in the frozen initial potential,
computed with the multigrid potential solver. No gas dissipation or star formation
are included in this relaxation phase.

\section{Results}
\label{res}
  To isolate the main physical parameters, purely stellar models 
are run first, although they do not correspond to the reality of galaxies in 
the early universe.  They are compared to analogous simulations,
where half of the baryonic mass is gas, which emphasizes the role of dissipation,
star formation, and feedback. 

%
\begin{center}
\begin{table}[ht]
      \caption[]{Estimate of decay time scales for Newton+dark matter models} 
\label{tab:decay}
\begin{tabular}{lrr}
\hline
Model & 5 clumps$^{(1)}$ & 3 clumps$^{(2)}$  \\
\hline
dwarf & 1.4$\pm$0.2    & 0.7$\pm$0.12 \\
giant  & 0.5$\pm$0.07  & 0.2$\pm$0.04 \\
\hline
\end{tabular}
\\ $^{(1)}$ Each clump has a mass of 5\% of the baryonic mass
\\ $^{(2)}$ Each clump has a mass of 10\% of the baryonic mass
\\ Decay time scales are in Gyr
\end{table}
\end{center}

\subsection{Dissipationless models}
\label{stars}

  Two extreme mass models, dwarf and giant galaxies, were run with the 
same characteristic scales, but with masses that are different by an order of magnitude
(see Table \ref{tab:simul}). To quantify the effects of dynamical friction,
initial clumps were simulated by massive particles, in two options:
either five clumps in all containing 25\% of the disk mass or
three clumps containing 30\% of the disk mass were distributed randomly
with the same initial velocity distribution as the rest of particles.

In total, eight types of runs were carried out with two mass models, with two types of clumps,
and for the two gravity models. For each type of run, ten different random realizations
of the initial conditions were used, to average out the results.
The decay of the clumps through dynamical friction is illustrated
in some of these runs in Figure \ref{fig:decay}.
 An estimate of the decay time scale (time for the clumps to fall
from their initial radius to the central region at 1-2kpc where their
orbit reaches a stationary state) 
is obtained through averaging of these clump orbits,
and is gathered in Table \ref{tab:decay} for the Newton+DM runs. 
All realizations of a given run type
are averaged out, with all orbits that are selected randomly. This procedure
looks more realistic than launching a given clump alone at an external radius,
since the dynamical effects of all the clumps interfere in the deformations
they impact on the dark halo and other particles in the disk.
This interference of several clumps can reduce and sometimes increase the dynamical friction effect
(Weinberg 1989). Also the decay time scales depend on the launching radius,
and in the same run, the clumps starting at smaller radii have a shorter
decay time scale.

Although these final averaged values have relative error bars of 10\% to 20\%, they are
compatible in their mass and density dependency with the Chandrasekhar
formula for dynamical friction (Chandrasekhar 1943). The decay time 
is predicted to be inversely proportional to the clump mass, for a 
given background distribution, and with the present assumptions on 
radii and mass ratios between clumps and disk, roughly inversely 
proportional to the square root of the total host mass. It is only 
an approximation of course, since many other parameters enter into account,
such as the initial velocity of the clump and the dark and baryonic radial mass
distributions.
It was not possible to measure the
corresponding values in the MOND models, because of weak dynamical friction.

\begin{figure}[ht]
\centerline{
\includegraphics[angle=-0,width=8cm]{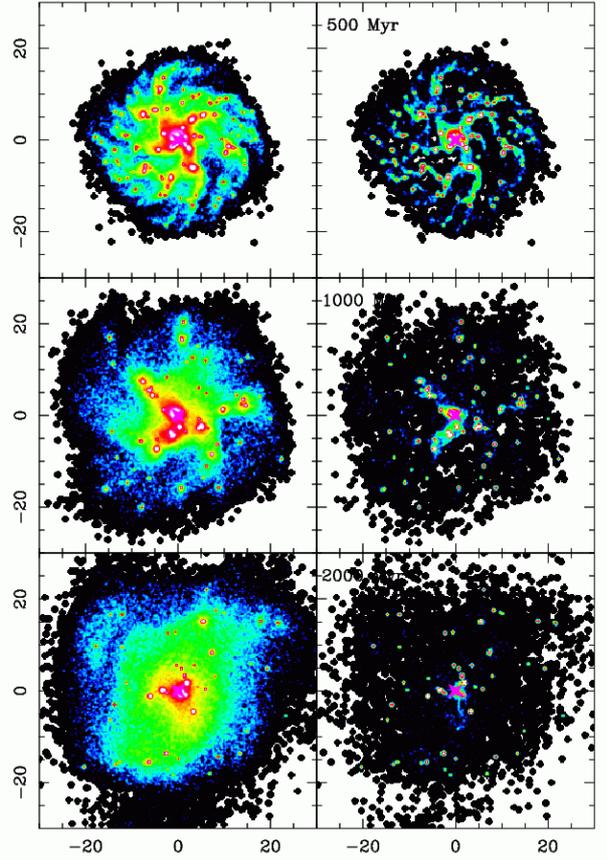}}
\caption{All baryons (left) and gas (right) surface densities of the dwarf
clumpy galaxy, simulated with MOND gravity, at epochs 0.5, 1, and 2 Gyr.
Each panel is 60 kpc in size. The color scale is logarithmic
and is the same for all plots.}
\label{fig:qclump}
\end{figure}

\begin{figure}[ht]
\centerline{
\includegraphics[angle=-0,width=8cm]{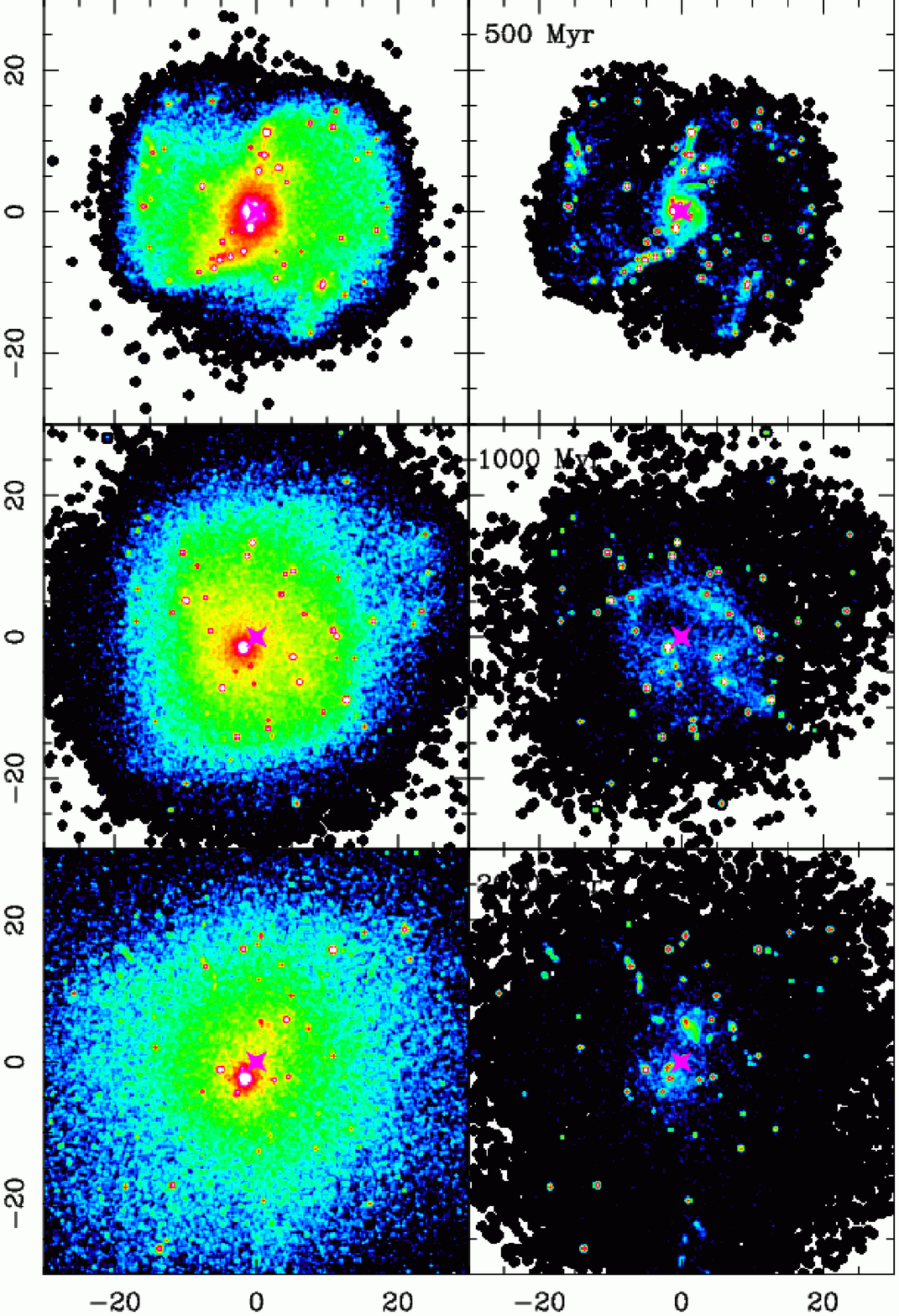}}
\caption{All baryons (left) and gas (right) surface densities of the giant
clumpy galaxy, simulated with MOND gravity, at epochs 0.5, 1, and 2 Gyr.
Each panel is 60 kpc  in size. The color scale is logarithmic
and is the same for all plots.}
\label{fig:qclump2}
\end{figure}

\begin{figure}[ht]
\centerline{
\includegraphics[angle=-90,width=8cm]{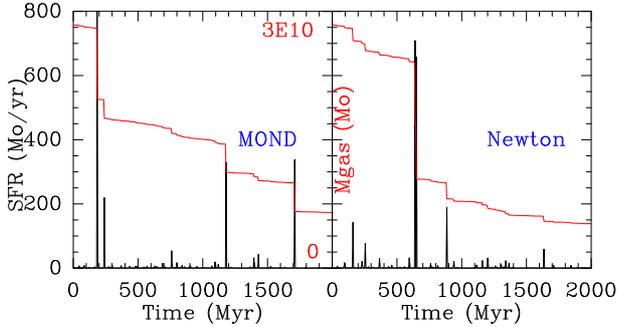}}
\caption{Evolution of the star formation rate (black line, left scale) and the
remaining amount of gas (red lines and middle scale) in the giant galaxy,
with MOND (left), and Newtonian dynamics (right). }
\label{fig:sfrate}
\end{figure}

\subsection{Models with gas, star formation, and feedback}
\label{gas}
\begin{figure}[ht]
\centerline{
\includegraphics[angle=-0,width=8cm]{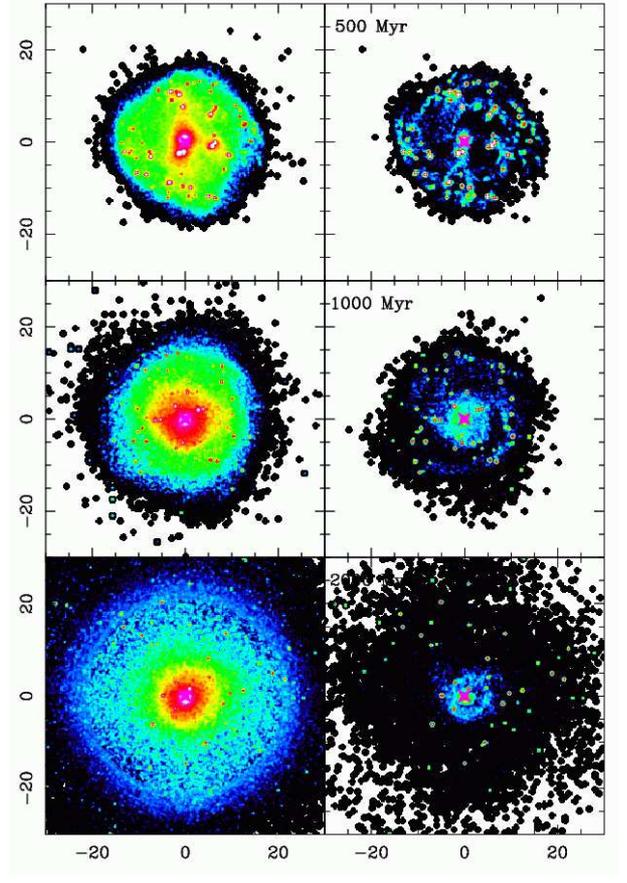}}
\caption{All baryons (left) and gas (right) surface densities of the giant
clumpy galaxy, simulated with Newtonian+DM gravity, at epochs 0.5, 1, and 2 Gyr.
Each panel is 60 kpc in size. The color scale is logarithmic
and the same for all plots.}
\label{fig:newtg2}
\end{figure}

Figures \ref{fig:qclump},  \ref{fig:qclump2}, and \ref{fig:newtg2}
show the evolution of the gas and total baryonic component of some
models with gas. The disks are highly unstable to clump formation
owing to the high gas fraction of 50\% and the absence of a bulge. Control runs 
with only stars were relatively stable, but they formed a strong bar,
after 500 Myr for the MOND disks and more than 2 Gyr for the Newtonian ones.
 The mass in clumps increases at the beginning with the gravitational 
instability due to the gas dissipation and large gas fraction. But these overdense
regions are very efficient at forming stars, and the gas fraction in clumps decreases,
as does the global gas fraction. Figure \ref{fig:sfrate} shows the evolution 
of the star formation rate (SFR) with time and the corresponding consumption of the gas.
 Since the SFR law adopted is nonlinear, the formation of dense clumps
leads to intense starbursts with an episodic character, which consume
the gas much more quickly than  the analogous quiescent, almost homogeneous disk galaxy,
which was used to calibrate depletion time.

 In the absence of external gas accretion and replenishment of the gas content, 
the disk becomes more
stable. Through both the star formation feedback and the gravitational shear
on the stellar clumps, the mass fraction in clumps slowly decreases.
The dynamical friction due to the dark halo on the clumps is, however, fast
enough to bring the clumps towards the center and build a bulge in the
Newtonian runs. In the MOND models, the stellar fragments are maintained in
the disk, because of  the weaker dynamical friction from the disk.

\begin{figure}[ht]
\centerline{
\includegraphics[angle=-90,width=8cm]{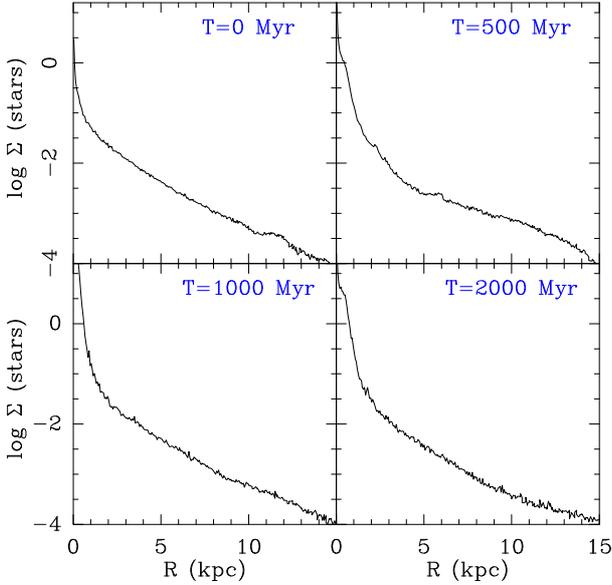}}
\caption{Stellar profile of the giant galaxy in the Newtonian dynamics,
at various epochs (0, 0.5, 1, and 2 Gyr). A bulge is building
alongside the exponential disk.}
\label{fig:prof-ens2}
\end{figure}

\begin{figure}[ht]
\centerline{
\includegraphics[angle=-0,width=8cm]{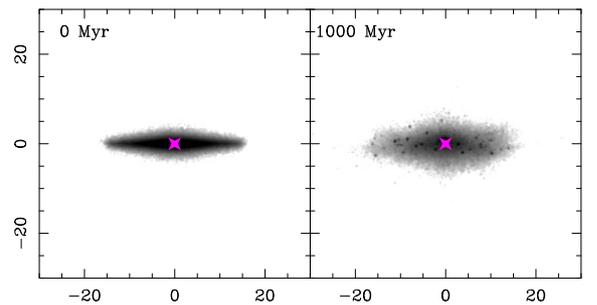}}
\caption{Evolution of the edge-on morphology of the giant galaxy,
in the Newtonian gravity model with gas and star formation.}
\label{fig:edge}
\end{figure}

The evolution of the stellar surface density $\Sigma$ as a function of time
is shown in Fig. \ref{fig:prof-ens2} for the giant galaxy in the Newtonian model
and in Fig. \ref{fig:prof-ens2} in the MOND gravity.
Initially, there is an exponential distribution in the disk, while a second component is
progressively accumulated toward the center in the
clumpy galaxy with dark matter. An edge-on view reveals that
there is indeed a small spheroid in the center, along with a thick disk (Fig. 
\ref{fig:edge}). This initial bulge formation due to massive clumps
has been discussed widely in the literature (Noguchi 1999, Immeli et al. 2004,
Bournaud et al 2007b, Elmegreen et al. 2008), along with the formation of a thick 
disk from star formation 
in a turbulent gas disk (Bournaud et al. 2009).
 We show that this rapid and early bulge formation does not occur
in the MOND gravity model. The disk remains clumpy until 2~Gyr, as revealed
by Fig. \ref{fig:prof-ens2}. This is the natural consequence of a much 
longer dynamical friction time scale.

\begin{figure}[ht]
\centerline{
\includegraphics[angle=-90,width=8cm]{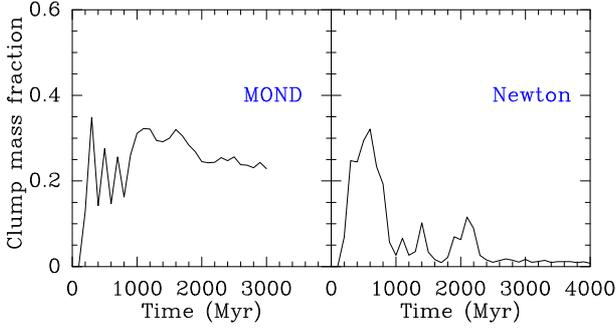}}
\caption{Evolution of the clump mass fraction (see text for the definition) for
the giant galaxy, in the MOND gravity (left) and 
in the Newtonian gravity (right).}
\label{fig:clumpf}
\end{figure}

The mass fraction in clumps was monitored as a function of time,
in the same manner as described by Bournaud et al. (2007b).  The surface
density of the baryonic component was computed at each snapshot. 
It was first smoothed at a resolution of 700 pc to reduce noise, and 
the azimutal average surface density $\Sigma$(r) computed at each radius.
When the local surface density was higher than three times $\Sigma$(r), it was considered
to belong to a clump. When plotted together, these regions do correspond
to the main clumps seen in the maps. There is no ambiguity with spiral arms,
since the clumps are dominating the global spiral structure (see Fig. \ref{fig:qclump}).
 The number of clumps found by this method is typically 10 to 20. The fraction of the disk 
mass in clumps can reach at the maximum of the starburst of the order of 30\%.
Figure \ref{fig:clumpf} displays the clump mass fraction evolution for the giant
model galaxy. It is remarkable that in the MOND regime, the disk galaxy remains 
highly clumpy, while in the Newtonian gravity with a massive dark matter halo,
the clumps are driven toward the center, and the disk becomes more homogeneous
after some last starbursts have consumed the remaining gas.

\begin{figure}[ht]
\centerline{
\includegraphics[angle=-90,width=8cm]{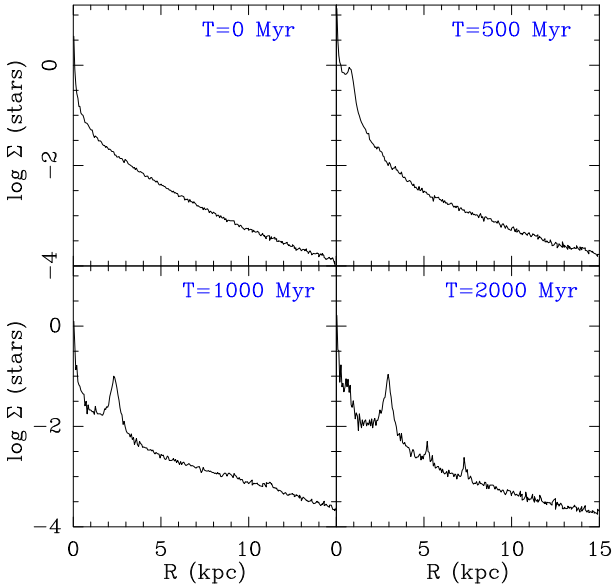}}
\caption{Same as Fig. \ref{fig:prof-ens2} for the MOND model.}
\label{fig:prof-qclump2}
\end{figure}

\subsection{Dynamical friction}
\label{DF}

The different dynamical friction efficiency in the two gravity models
is the key to understanding the morphology of clumpy galaxies in the early universe
and the bulge formation. The main phenomenon at the origin of the drag on the 
clumps is the deformation or wake
that a massive body imprints on the background distribution of particles. When the 
bodies have low mass with respect to the total system, their wake is a
small perturbation, and the process can be handled analytically. However,
with massive bodies, it is not an intuitive process to
derive the actual effect of the friction,
especially when the massive bodies are numerous, and their wakes interfere.
They can perturb the system significantly, so that it cannot be considered
as an infinite reservoir of energy and angular momentum. 

Results that may look controversial at first glance have been claimed in the literature.
A perturbation study by Ciotti \& Binney (2004) concludes that dynamical friction is stronger
in MOND. They consider only very low-mass
bodies, like star clusters, with a negligible effect on the full stellar
background, and adopt an impulse approximation
for deflection or orbits and linear summation of effects. For the Newtonian model with dark matter, they 
consider the dark matter halo as a rigid
background that does not participate in the dynamical friction. They find that
the two-body relaxation time is shorter for disks in the MOND regime
and extrapolate these results to the dynamical friction
time, obtained for test particles in the local formula of Chandrasekhar (1943).
The consequences would be that globular clusters should spiral inwards
to the center in dwarf galaxies on a few dynamical time scales,
as well as galaxies in groups and in clusters. Simulations
by Nipoti et al. (2008) confirm these findings in the context of
 tiny perturbations:
the massive bodies subject to the friction, either globular clusters or a
rigid bar, have to contain less than 5\% of the baryonic mass, so that particles
in the stellar background absorbing the energy and angular momentum are 
not significantly perturbed.
In realistic systems though, Nipoti et al. (2007) find that the merging time scales
 for spherical systems are significantly longer in MOND than in Newtonian gravity
with dark matter, and Tiret \& Combes (2007) found that bars keep their pattern speed
constant in MOND, while they are strongly slowed down in the Newtonian equivalent
system with a dark matter halo.

The main conclusion of these apparently different claims is that
the dynamical friction time scale is greater in MOND for satellite
galaxy interactions and for stellar bars, since
galaxies are not embedded in extended and massive spheroids of dark matter particles, which are
able to accept the orbital angular momentum. 
Although the impact of very small fluctuations could be greater in MOND
than in Newtonian dynamics, the effect saturates quickly when the perturbation
is no longer infinitesimal. In contrast, the equivalent Newtonian system
with dark matter has shorter dynamical time scales, and common satellites and bars
are slowed down in a few rotation times.
The present results extend these conclusions for clumpy galaxies,
which are typical of high-redshift systems, where the total clump mass fraction
can reach 30\% of the disk mass, with individual clumps of 10$^8$-10$^9$\msol.
The typical time scale for the clumps to be driven to the center and form
the bulge in the Newtonian model with dark matter
is about 1~Gyr for the giant galaxy and 3~Gyr for the dwarf.
For the MOND model, the dynamical friction is not efficient enough to form
a bulge before the clumps are destroyed or reduced by the star formation feedback
and the tidal and shear forces.

\section{Discussion}
\label{disc}

The present simulations are based on the simple model of sticky particles for the gas physics
and dissipative character. Previous works have used different methods, either sticky particles
(Noguchi 1999, Immeli et al 2004, Bournaud et al 2007a, Elmegreen et al 2008), AMR
hydro-code (Dekel et al 2009, Ceverino et al. 2010, Behrendt et al. 2014) ,
or Tree-SPH code (Hopkins et al. 2012).
The results are very similar, as far as the instability of the early galaxies is concerned.
 Massive clumps are formed, reproducing clumpy galaxies at high redshift. The importance
of the bulge formation is variable according to the stellar feedback adopted. Clumps
are driven to the center through dynamical friction in less than 1 Gyr for massive
galaxies of $\sim$ 7 10$^{10}$ \msol\, in baryons, but extreme supernovae feedback could 
significantly reduce the mass of clumps ending in the bulge (Hopkins et al. 2012). 
The bulge in all cases has a stabilizing influence on the disk, and after the clumpy phase,
the disk remains quite homogeneous, even in the case of abundant external gas accretion
(Ceverino et al 2010).
We have tried several degrees of dissipation, and the variation
of the collisional parameters $\beta$ between 0.3 and 0.7 did not have a strong influence on 
the results, confirming the conclusions of Bournaud et al. (2007a).
The essential point is to have sufficient dissipation in the gas to allow its collapse
into dense clumps and to enhance the star formation rate through a nonlinear 
Schmidt law.

\begin{figure}[ht]
\centerline{
\includegraphics[angle=-90,width=8cm]{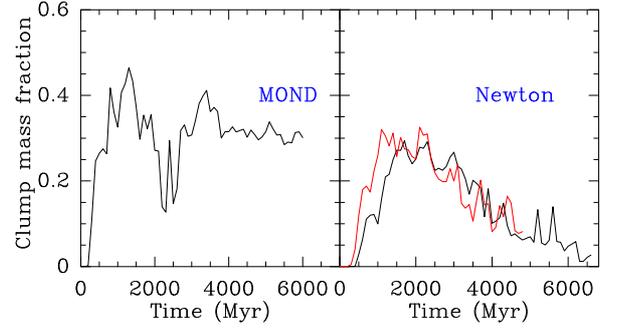}}
\caption{Evolution of the clump mass fraction for
the dwarf galaxy in the MOND gravity (left) and 
in the Newtonian gravity (right). The high-resolution run with a 513$^3$ grid
and 8 times the number of particles is plotted in red.}
\label{fig:clumpf-dwarf}
\end{figure}

It is also necessary to check that the spatial resolution of the simulations
is not perturbing the main conclusions. The 257$^3$ grid means a resolution of 230~pc
for the potential solver, it is always lower than the Jeans scales.
From initial conditions, it is possible to predict the range in sizes and 
masses of clumps in the unstable disks (but see Behrendt et al. 2014). 
The Jeans length in an axisymmetric disk is $\lambda_{J}= \sigma_g^2/(G \Sigma)$, 
where $\sigma_g$ is the gas velocity dispersion, and 
 $\Sigma$  the disk's surface density.  These range from 0.5 to 1~kpc in the dwarf
galaxy model and from 1.4 to 2~kpc for the giant. The corresponding ranges in mass
of the clumps are M$_{J}= \pi \sigma_g^4/(G^2 \Sigma)$, 
between 10$^6$-10$^8$\msol\, for the dwarf galaxy, and
10$^8$-10$^9$\msol\, for the giant model.
A higher resolution simulation with 513$^3$ grid has been run for the dwarf galaxy
model, increasing the number of particles to six million. The clump mass fraction
is plotted in Fig. \ref{fig:clumpf-dwarf}. In general, the evolution is similar, 
with the clump fraction determined with the smoothing length adapted
to a factor-2 higher spatial resolution.

Another interesting issue is the stability of the considered disks to 
large-scale waves, although small-scale fragmentation is dominant.
The critical wavelength $\lambda_{crit}$ above which disks should be stabilized
by differential rotation, even without pressure (dispersion) terms, 
indicating that the self-gravity scale is 
$\lambda_{crit} = 4 \pi^2 G \Sigma /\kappa$, where $\kappa$ is the epicyclic frequency. It ranges from 1-3~kpc in the dwarf galaxy and from 2-6~kpc for the giant. The role of 
the dark matter halo is proportionally much higher in the dwarf.
The X-Toomre parameter, X = $\lambda/\lambda_{crit}$, where $\lambda\sim 2\pi r$
controls the self-gravitating response of the disk through
the swing amplifier; waves develop more efficiently when X$\sim$ 2
(Toomre 1981). In the present galaxy models, a 
central region always exists where X $\sim$ 1-2.

\section{Summary}
\label{summary}

Simulations have been carried out of the dynamical processes occurring
early in the universe, in galaxies that are very rich in gas, and corresponding
to the clumpy galaxies observed at high redshift (e.g.,
Elmegreen \& Elmegreen 2005, Elmegreen 2007). Previous works have established that
bulge formation is rapid in those galaxies through dynamical friction driving
the clumps to coalesce in the galaxy centers (Noguchi 1999, Bournaud et al. 2007b,
Elmegreen et al 2008). The present work explores the analogous phenomena
in modified gravity (MOND, Milgrom 1983). A multigrid code is used to solve
the nonlinear equations, both from the original AQUAL Lagrangian (Bekenstein
\& Milgrom 1984) and the QUMOND solution, recently developed by Milgrom (2010).
 Both appear to give similar results, and most runs were performed in the
faster method of QUMOND.
For each galaxy model, the corresponding rotation curve is first computed in
MOND, and then an equivalent Newtonian system is built with a dark matter
halo added, in order to get exactly the same rotation curve as in
Tiret \& Combes (2007, 2008a).
The efficiency of dynamical friction was first compared between the two gravity models,
with academic purely stellar runs, where an ensemble of clumps were
randomly distributed initially with 20-30\% of the baryonic mass.
 A series of such simulations show that clumps should spiral inwards to the center
in the Newtonian galaxies in about 1~Gyr or less, while the clumps remain orbiting
in the disk for more than 2-3 Gyr in the MOND regime.

The most realistic simulations include gas dissipation, star formation, and feedback.
Disks are initially composed of 50\% stars and 50\% gas.
Rapid fragmentation occurs in these unstable disks, and clumpy galaxies that are
similar to what is observed at high redshift are formed in both gravity models.
This clumpy morphology lasts less than 1~Gyr in the giant galaxy model
of baryonic mass 5.7 10$^{10}$ \msol and 3 Gyr in the dwarf galaxy, with
ten times less mass. Clumps are driven inwards by
dynamical friction, and form a spheroidal bulge.
This is very compatible with previous works.
However, in the MOND gravity model, the clumps remain in the disk for
a much longer time, are eventually destroyed or reduced by stellar
feedback and shear, and have no time to fall in the center to form a bulge.
 The lower bulge formation efficiency in MOND is more in line with
the large number of observed bulgeless galaxies today (Kormendy et al. 2010).
 
Fisher \& Drory (2011) also find that the galaxies that contain either a pseudo-bulge or no bulge
represent more than 80\% of galaxies above a stellar mass of 10$^9$\msol\, in the 
local volume of 11 Mpc radius. 
This is difficult to explain in the standard Newton plus dark matter models.
Weinzirl et al. (2009) compare the observations of the bulge mass and frequency in present-day
galaxies with the predictions of the CDM model. Massive galaxies can have less than 20\%
of their baryonic mass in a bulge if they experience mergers only before z=4. But
 galaxies having later mergers and larger fractions of their mass in a bulge are
 about 30 times more abundant in CDM simulations than observed.
Zavala et al. (2012) and Avila-Reese et al (2014) revisit the CDM predictions, with
semi-enpirical recipes and an analytical treatment to follow the transfer of stellar
masses to bulges after a merger. They assign the stars coming from merging satellites to 
the classical bulges and stars coming from the primary during the merger to pseudo-bulges.
But they do not consider the pseudo-bulges formed by secular evolution of bars.
They then claim to form less classical bulges, since there is now only one path to form them.
Also the result depends on the adopted evolution of the stellar-to-halo mass relation, the mass assembly 
in their models resembling more a monolithic scenario than a hierarchical one.
In any case, there are still problems explaining the large fraction of bulgeless galaxies
observed locally.

In the framework of MOND, classical bulges are hardly formed in early times in the 
clumpy phase of galaxy formation, when gas is dominating the surface density of galaxy disks.
They can form later through hierarchical merging with a frequency that is
smaller than what occurs in the analogous Newtonian systems with dark matter.
Pseudo-bulges, however, form with equivalent frequency through secular evolution
by vertical resonances with bars.  It is therefore expected that the relative frequency 
of pseudo-bulges is higher in MOND. More quantitative statements will require
many more simulations with a cosmological context. 

\begin{acknowledgements}
 I warmly thank the referee, Stacy McGaugh, for constructive comments and suggestions. 
The idea of this work came when Eija Laurikainen asked me to write a review article on
bulge formation with MOND.
The European Research Council
for the Advanced Grant Program Number 267399-Momentum
is acknowledged. Most simulations have been run on the cluster
provided by this ERC grant. 
\end{acknowledgements}


\begin{thebibliography}{}
\bibitem{}Avila-Reese V., Zavala J., Lacerna I.: 2014, MNRAS 441, 417
\bibitem{}Barnes, J. E.: 1988, ApJ 331, 699
\bibitem{}Barnes, J. E., Hernquist, L. E. 1991, ApJ 370, L65
\bibitem{}Behrendt, M., Burkert, A., Schartmann, M.: 2014, MNRAS sub (arXiv1408.5902)
\bibitem{}Bekenstein J., Milgrom M.: 1984, ApJ  286, 7
\bibitem{}Bertone G., Hooper D., Silk J: 2005, PhR 405, 279
\bibitem{}Blumenthal G.R., Faber S.M., Primack, J.R., Rees M.J.: 1984, Nature 311, 517
\bibitem{}Bode P., Ostriker J.P., Turok N.: 2001, ApJ 556, 93
\bibitem{}Bournaud F., Jog C.J., Combes F.: 2005, A\&A 437, 69
\bibitem{}Bournaud F., Jog C.J., Combes F.: 2007a, A\&A 476, 1179
\bibitem{}Bournaud F., Elmegreen B.G., Elmegreen D.M.: 2007b ApJ 670, 237
\bibitem{}Bournaud F., Elmegreen B.G., Martig M.: 2009, ApJ 707, L1
\bibitem{}Boylan-Kolchin, M., Bullock, J. S., Kaplinghat, M.: 2011, MNRAS 415, L40
\bibitem{}Boylan-Kolchin, M., Bullock, J. S., Kaplinghat, M.: 2012, MNRAS 422, 1203
\bibitem{}Brada R., Milgrom M.: 1999, ApJ 519, 590
\bibitem{}Ceverino D., Dekel A., Bournaud F.: 2010, MNRAS 404, 2151
\bibitem{}Chandrasekhar S., 1943, ApJ 97, 255
\bibitem{}Ciotti L., Binney J.: 2004, MNRAS 351, 285
\bibitem{}Combes F., Debbasch F., Friedli D., Pfenniger D.: 1990, A\&A 233, 82
\bibitem{}Dekel A., Sari R., Ceverino D.: 2009, ApJ 703, 785
\bibitem{}Elmegreen B.G., Elmegreen D.M.: 2005, ApJ 627, 632
\bibitem{}Elmegreen D.M.: 2007, in IAU S235, ed. F. Combes \& J. Palous, CUP, p. 376
\bibitem{}Elmegreen B.G., Bournaud F., Elmegreen D.M.: 2008, ApJ 688, 67
\bibitem{}Famaey B., Binney J.: 2005, MNRAS 363, 603
\bibitem{}Famaey B., McGaugh S. S.: 2012, Living Reviews in Relativity, vol. 15, no. 10
\bibitem{}Fisher D.B., Drory N.: 2011, ApJ 733, L47
\bibitem{}Hopkins, P. F., Keres, D., Murray, N. et al.: 2012, MNRAS 427, 968
\bibitem{}Immeli A., Samland M., Gerhard O., Westera P.: 2004, A\&A  413, 547
\bibitem{}Jungwiert B., Combes F., Palous J.: 2001, A\&A 376, 85
\bibitem{}Kennedy R., Frenk C., Cole S., Benson A.: 2014, MNRAS 442, 2487
\bibitem{}Kennicutt R.C.: 1998, ApJ 498, 541
\bibitem{}Kormendy, J., Kennicutt, R. C.: 2004, ARAA 42, 603
\bibitem{}Kormendy J., Drory N., Bender R., Cornell M. E., 2010, ApJ 723, 54
\bibitem{}L\"ughausen F., Famaey B., Kroupa P. et al.: 2013, MNRAS 432, 2846 
\bibitem{}Milgrom M.: 1983, ApJ 270, 365
\bibitem{}Milgrom M.: 2010 MNRAS 403, 886 
\bibitem{}Naab, T., Burkert, A. 2003, ApJ 597, 893
\bibitem{}Nipoti, C., Londrillo, P., Ciotti, L., 2007, MNRAS 381, 104
\bibitem{}Nipoti, C., Ciotti, L., Binney, J., Londrillo, P.: 2008, MNRAS 386, 2194
\bibitem{}Noguchi M.: 1999, ApJ 514, 77
\bibitem{}Peebles P.J.E., Ratra B.: 2003, RvMP  75, 559
\bibitem{}Press W.H., Teukolsky S.A., Vetterling W.T., Flannery B.P.,
Numerical Recipes in Fortran 77, second edition, Cambridge
University Press, 1992
\bibitem{}Sanders R.H., McGaugh S.: 2002, ARAA 40, 263
\bibitem{}Tiret O., Combes F.: 2007, A\&A 464, 517
\bibitem{}Tiret O., Combes F.: 2008a, A\&A 483, 719
\bibitem{}Tiret O., Combes F.: 2008b ASPC 396 , 259
\bibitem{}Tiret O., Combes F.: 2009, A\&A 496, 659
\bibitem{}Toomre A.: 1977, in Evolution of Galaxies and Stellar Populations, Proceedings of a Conference at Yale University, p. 401
\bibitem{}Toomre A.: 1981, Proceedings of the Advanced Study Institute, Cambridge, England, CUP  p.111-136
\bibitem{}Wang J., White S.D.M.: 2009, MNRAS 396, 709
\bibitem{}Weinberg, M.D.: 1989, MNRAS 239, 549
\bibitem{}Weinzirl T., Jogee S., Khochfar S. et al.: 2009, ApJ 696, 411
\bibitem{}Zavala J., Avila-Reese V., Firmani C., Boylan-Kolchin M.: 2012, MNRAS 427, 1503
\bibitem{}Zhao, H.S., Famaey, B.: 2006, ApJ 638, L9
\bibitem{}Zhao, H.S., Famaey, B.: 2010, Phys. Rev. D, 81(8), 087304
\end{thebibliography}
\end{document}